# Morphogenesis Software based on Epigenetic Code Concept

N. Bessonov[1], O.Butuzova[2], A.Minarsky[3], R. Penner[4], C. Soulé[4], A. Tosenberger[5] and N. Morozova[2,4,6*]

[1]IPME RAS, St-Petersburg, Russia; [2]BIN RAS, St-Petersburg, Russia; [3]Academic University, St-Petersburg, Russia, [4]IHES, France; [5] ULB, Belgium; [6]UMR9198, I2BC, CNRS, France;

\* corresponding author

**Abstract**

The process of morphogenesis is an evolution of the shape of an organism together with the differentiation of its parts. This process encompasses numerous biological processes ranging from embryogenesis to regeneration following crisis such as amputation or transplantation. A fundamental theoretical question is where exactly do these instructions for (re-)construction reside and how are they implemented?

We have recently proposed a set of concepts, aiming to respond to these questions and to provide an appropriate mathematical formalization of the geometry of morphogenesis [1]. First, we consider the possibility that evolution of shape is determined by epigenetic information, responsible for realization of different types of cell events. Second, we suggest a set of rules for converting this epigenetic information into instructive signals for cell events for each cell, as well as for transforming it after each cell event. Next we give notions of *cell state*, determined by its epigenetic array*,* and *cell event,* which is a change of *cell state*, and formalize development as a graph (tree) of cell states connected by 5 types of cell events, corresponding to the processes of cell division, cell growth, cell death, cell movement and cell differentiation.

Here we present a Morphogenesis Software capable of simulating the evolution of a 3D embryo starting from zygote, following a set of rules based on our theoretical assumptions, and thus to provide a proof-of-concept for the hypothesis of epigenetic code regulation. The software creates a developing embryo and a corresponding graph of cell events according to the zygotic epigenetic spectrum and chosen parameters of the developmental rules. Variation of rules influencing the resulting shape of an embryo may help elucidating the principal laws underlying pattern formation.

# 1. Introduction

The process of morphogenesis is an evolution of a shape of an organism together with the differentiation of its parts. The discovery of differential gene expression (the spatial–temporal distribution of the gene expression pattern during morphogenesis), together with its key regulators (such as Hox genes), is one of the main recent achievements in developmental biology. Nevertheless, solely differential gene expression cannot explain the development of the precise geometry of an organism and its parts in space.

Here we aim to model conceptual laws, underlying creation of determined morphology (geometry) of organisms, and not take into consideration the mechanisms, implementing these laws, such as morphogen gradients, electrical and mechanical signaling and differential gene expression. These important components of morphogenesis are well described for many concrete developmental cases, being reflected in numerous mathematical models. Numerous recent works are devoted to mechanisms of morphogen regulation [2-7] and corresponding models [8-10], to electrical signaling in morphogenesis [11-14] and to the role of mechanical cues [15-18]. However, the conceptual gap between a set of particular mechanisms and creation of a concrete morphology is still not filled, thus representing an intriguing field both theoretical [19-22] and computational [23, 10].

As a possible response to this problem, we conjecture the existence of an additional biological code (epigenetic code) which bears information about geometrical pattern of an organism and thus coordinates the cascades of molecular events implementing a pattern formation (e.g., differential gene expression, directed protein traffic, growth of cytoskeleton). We understand the term epigenetic in a broad sense, as any information in a cell additional to the genetic one, which can be inherited by cells and be involved in regulation of their cell fates in a tight interplay with a genetic code. By that, we can consider a wide spectrum of possible levels of epigenetic information. We assume that a vast set of all intracellular processes, important for morphogenesis, is governed and controlled by such an epigenetic code; thus, we do not aim to model each of them, but rather to provide a framework for exploring a conceptual epigenetic control theory.

Though the concrete signal transduction pathways connecting the morphogenetic information and expression of given sets of genes are not yet elucidated, we can suggest a set of postulates and possible approaches for discovering the correspondence between epigenetic code and its realization in a given geometry of an organism in space–time.

## 2. Epigenetic Code Concept

The set of theoretical conjectures on the geometry of morphogenesis built on the hypothesis of an existence of epigenetic cell surface code is a further development of work published in [1, 24, 25].

First we suggest as a model that a cell fate and, correspondingly, a final pattern of a multicellular object, is coded by a biological code located on cell surfaces in such a way that with each cell can be associated a corresponding matrix/coding array, reflecting a 3D pattern of distribution of a set of coding molecules on cell surfaces.

Next we suggest that a set of rules (*developmental laws*) for converting this coded information into instructive signals for cell events for a cell, as well as for transforming the coding arrays after each type of cell event, may be common to all living organisms.

In this case, development of an organism depends on a coding array of its initial cell, its zygote. Next after each cell division, daughter cells inherit a part of a *coding array* of the mother cell, thus providing a basis for differential developmental paths of cells containing the same DNA content.

We provide a set of arguments why coding molecules should be located on the cell surface, and suggest a set of experiments for the confirmation of this concept. Concerning the molecular character of coding molecules, our prevailing **assumption** is that such code may be provided by a pattern formed by a set of several types of oligosaccharide residues of glycoconjugates (glycoproteins and glycolipids), some specific features of which make them plausible candidates. However, in the full context of the general model, we can also consider any type of cell surface markers.

Next we give notions of *cell state* which is determined by its *coding matrix,* and *cell event,* which is a change of *cell state*, and formalize development as a graph (tree) of cell states connected by 5 types of cell events, corresponding to the processes of cell division, cell growth, death, cell movement and cell differentiation. We show that such a developmental tree with exact parameters of cell events has one-to-one correspondence with an embryo morphology at each time slice.

Next, we assume that there is a universal rule $R_i$ for changes in the amount and composition of cell surface markers for each type $i$ of *cell event*.

We suggest a mathematical formalism suitable to decipher these rules (*developmental laws)* for converting the coded morphogenetic information into instructive signals for cell events for a cell, and a corresponding software which gives a tool for determination of cell events based on the distribution of epigenetic code and the rules of epigenetic code change following cell events.

**Hypotheses and conjectures.**

1. Existence of morphogenetic code, determining a geometrical outcome of developing organism, located on cell surfaces. The arguments why coding molecules should be located on the cell surface are:

- The points of a cell surface correspond to the geometrical structure in 3D space thus giving an intrinsic metric which can be used for recording spatial information
- A cell surface location provides the possibility of different distributions of this information within a set of dividing cells of an organism, hence providing a diversity of cell potencies for further differentiation
- This location gives a possibility of feedback, software, the instruction to stop, when the task for a proper shape (for a cell, morphological domain, or a whole organism) is fulfilled
- This location gives a possibility to be involved in the signal transduction pathways. For example, received outer signals go to the nucleus or Golgi apparatus and influence the expression of specific set of genes or the process of protein traffic
- This location gives a possibility to influence a direct cell-cell communication
- The cell surface of an ovule is inherited as well as its DNA content

- A set of experimental data confirms the significance of cell surface information for pattern formation in animals (summarized in [26]) and plants [27-29]). As an additional example of the importance of cell surface information, a behavior of protoplasts (plant cells without a cell wall, removed by specific enzyme) in cell culture can be considered. As it is described in classical plant cell culture methodology (e.g., [30]), protoplasts, produced from terminally differentiated cells (of leaves, fruits, etc.) upon removing a cell wall start to proliferate, dividing eternally and producing a callus (an unorganized mass of undifferentiated cells), thus losing all preceding morphogenetic information.

2. Cell fate may be coded by a biological code located on cell surfaces in such a way that with each cell can be associated a corresponding matrix/array, containing this code

3. *Cell events of a cell* and its parameters implementing embryo development are functions of its coding matrix

4. There is a universal rule $R_i$ for changes in the amount and composition of surface markers for each type $i$ of *cell event* (cell division, cell growth, death, cell movement), which may be the same for all living organisms

5. After each *cell division* event daughter cells inherit one part of a *coding matrix* of the mother cell, while the remaining part is created according to a rule $R_d$

6. The set of rules (*developmental laws*) for converting coded information into instructive signals for cell events as well as for transforming the coding arrays after each cell event, may be common for all living organisms.

We propose a model implementing these ideas and hope to find the general rules of a surface code coordination of the pattern formation by mathematical formalization. We do not pretend for now to suggest any molecular mechanisms underlying this coordination.

## 3. Morphogenesis Software

### 3.1. Mathematical formalization of epigenetic code hypothesis

For checking this hypothesis by numerical simulations, we create a Morphogenesis Software, allowing the creation of a first cell (zygote) with a given coding matrix, which will develop into an embryo according to this set of rules (https://github.com/nickbessonov/morphogenesis-article/blob/master/Morphogenesis.zip). The program has an advanced interface that allows one to observe the process of growth of an embryo from a zygote simultaneously with a corresponding developmental graph (Figure 1), and to show the matrix of any selected cell during simulations. The program starts with a matrix of the first cell of an organism (zygote), which can be varied by the (computational) experimentalist and includes a set of rules for embryo development depending on cell matrix information. By applying these rules to all cells at each time step, the program presents a developmental process as a programmed consequence of *Cell events*, occurring at each time interval, and compute and displays both geometrical structure of a developing embryo and a corresponding graph (tree) of *Cell events* (**Figure 1**). In a graph, each horizontal layer corresponds to a *level* of cell development, where each new level appears after one *step*, which is a time period during which at least one cell event has occurred at least in one cell.

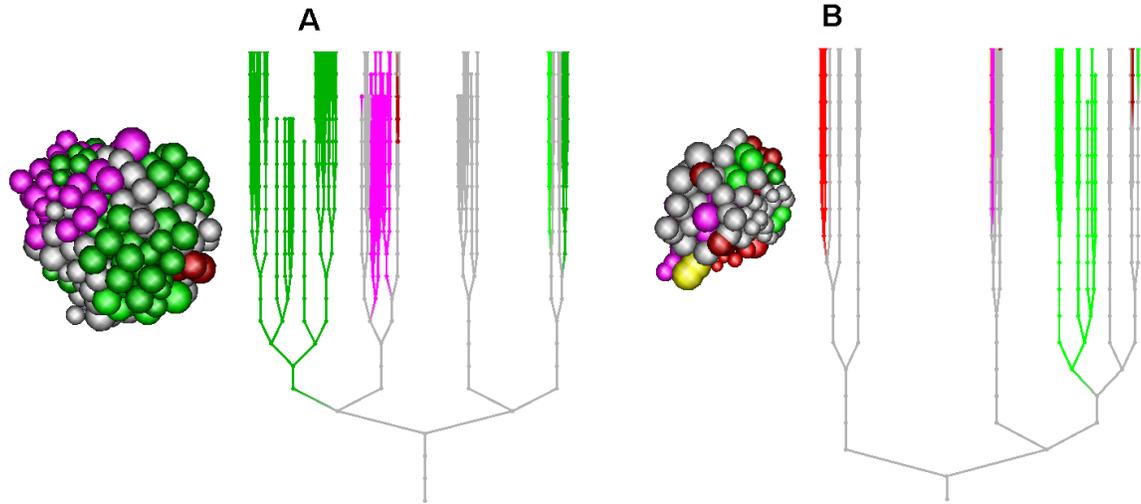

**Figure 1.** Developing embryo and corresponding Tree of Cell events. Different colors correspond to different types of differentiated cell.

This means that having a determined set of Rules, the main parameters of *Cell events* implementing embryo development are the functions of a code, located on a cell surface (a corresponding matrix associated with each cell). We consider 5 types of *Cell events*, determined by the algorithm: cell division, cell growth (including changing a cell shape), movement, death and differentiation. Also the algorithm includes the rules for filling in new elements of a coding matrix after each type of cell event (one rule for each type of cell event).

We consider that a cell surface code can be written and transmitted in the form of a matrix $a_{in}$, $i = 1, ..., I, n = 1, ..., N$ which has the following structure: $N$ columns of the matrix $a_{in}$ corresponds to $N$ sectors on a cell surface, while each row $i$ corresponds to one type of coding molecules (substances) $I$. An element of a matrix $a_{in}$ shows an amount of a given type of coding molecules $i$ in a sector $n$ of a cell surface, presented by an integer number.

Though it does not influence the calculations in the framework of this model for now, we would like to provide a suggestion for possible candidates for coding molecules. We postulate that there can be different types of oligosaccharides - short sugar residues of cell-surface glycoconjugates, i.e., proteins or lipids with sugar (glycol-) part. Oligosaccharides can be monosaccharides (mannose, glucose, galactose, rhamnose, fucose, xylose, etc.), or di- or tri-saccharides, combined from 2 or 3 monosaccharides.

In a simplest model we consider 8 sections of cell surface and 8 coding molecules (e.g., monosaccharide residues only), as shown on the software panel (Figure 2A). The spatial orientation of the sectors are numbered as a $2 \times 2 \times 2$ matrix with elements $A_{xyz}$, $(x = 0,1,\ y = 0,1, z = 0,1)$ (Figure 2B). In this case we can consider a vector $M$ with $I$ coding substances $M_i$, $(i = 1, ..., I)$ corresponding to each sector, and we consider $I = 8$ (marked as A, G, F, K, P, D, X, Y on the software panel).

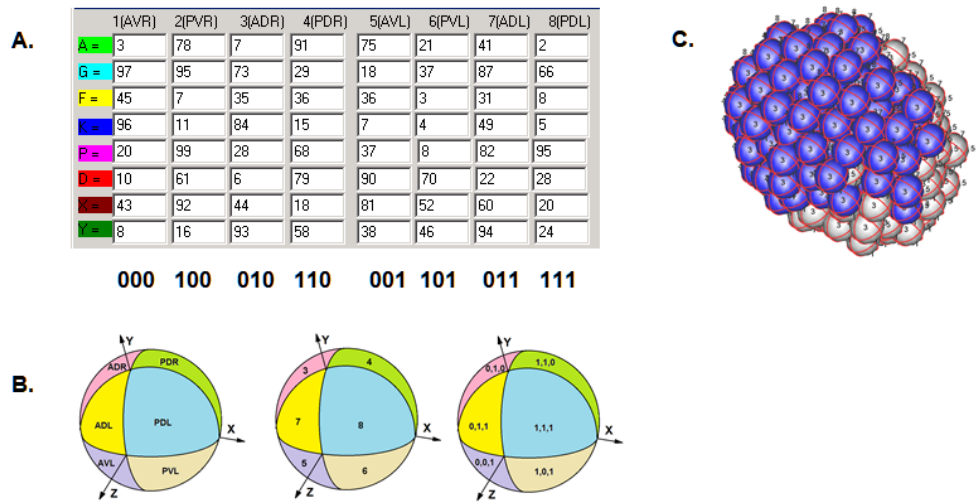

**Figure 2. A coding matrix, associated with a cell**. **A** - A coding matrix, shown in the panel of Morphogenesis Software. For each of the eight sectors of a cell the amount of 8 coding molecules A, G, F, K, P, D, X, Y in this sector are given as integer numbers. **B** - The visualization of spatial localization of 8 sectors (columns of the matrix) on a cell surface in a coordinate system, associated with a zygote: $xyz, x = 0,1, y = 0,1, z = 0,1$, where xyz corresponds to embryo axis AP, DV, LR. The corresponding spatial 3D coordinates are marked below each colon. **C**- an embryo with the sectors numbers marked on the surfaces of the cells.

The algorithm includes two types of interactions between cells, namely, the adhesion and the signaling, both determined by coding matrixes, associated with contacting cells. Later we are planning to add also a long-distance signaling, determined by secreted factors, produced by some of the cells and influencing cell events of other cells in the area of their effective concentrations.

Though we are currently working out an appropriate mathematical formalism which will allow deciphering of possible rules determining *cell events* as operators acting on a corresponding matrix of a *cell state*, for the time being we include in the algorithm of Morphogenesis Software a set of simple but biologically relevant rules for the proof-of-concept of the cell surface code hypothesis.

The detailed description of the Morphogenesis Software interface is presented in Supplemental Materials.

**3.2. Model**

According to our theory and formalization, we propose a set of rules for cell events implementing embryo development. Each rule reflects the dependence of a cell event on an epigenetic matrix of a cell, and at the same time is based on the biological nature of a concrete cell event. The function determining cell event can depend upon matrix structure (spatial distribution of coding molecules), so upon its general characteristics (e.g., its homogeneity, the percentage of zero elements in it, etc.).

**Rule 1. Choice of cell event depending on a cell coding matrix.**

We will consider a constant time period (step) corresponding to the duration of each cell event. The pipeline for the choice of a cell event, which takes place at the end of each step and occurs corresponding to a coding matrix of a cell is as follows.

First, for each new cell we check if the conditions for cell event *Apoptosis* (see below in Rule 6) are applicable. If yes, the cell undergoes the cell event *Apoptosis*. If no, we check the possibility of the cell event *Division*. For that, we consider each set of 2 halves of cell surface which can be obtained during cell division in 3 possible division planes (x,y,z), namely, the "left" and "right" halves $X_L$ and $X_R$ for the division plane x, $Y_L$ and $Y_R$ halves for the division plane y, $Z_L$ and $Z_R$ halves for the division plane z.

Next we calculate a sum of all components (A,G,F,K,P,D,X,Y) for each half and determine the Moment M of a cell for each axis:

$$\begin{array}{ll} Mx = |X_R - X_L| & \\ My = |Y_R - Y_L| \quad \text{and} & M_{min} = min(M_x, M_y, M_z) \\ Mz = |Z_R - Z_L| & M_{max} = max(M_x, M_y, M_z) \end{array}.$$

We will consider Conditions 1 and 2:

Condition 1:

$$\mathrm{M}_{max} - \mathrm{M}_{min} > \mathrm{P} \cdot m_h \cdot \lambda, \tag{1}$$

where P is a parameter of heterogeneity of zygotic matrix, showing the maximal dispersion of the values of its elements as a percentage in [0,1]; $m_h$ is the highest possible amount of coding molecules in the zygotic matrix (both are introduced as parameters), and the chosen constant parameter $0 < \lambda < 1$ is a parameter of inheritance (described in Parameters section).

Condition 2:

Both standard deviations $\sigma_{left}$ and $\sigma_{right}$ of all elements in both left and right halves of the surface related to the $M_{max}$ direction satisfy the condition

$$\sigma > e, \ e = X \cdot P \cdot m_h, \qquad (2)$$

where *e* determines a sensitivity of rules to a variance of the values of matrix elements, reflecting a sensitivity of biological mechanisms, e.g., signal transduction pathways governed by epigenetic code, to the local change of the amount of a particular type of coding molecule. The chosen parameter of sensitivity X in (2) should be taken in the interval [0,1], and suggested to be around 0,1.

If both conditions 1 and 2 are fulfilled, a cell will undergo the cell event *Division,* The division plane is determined by $M_{max} = \max(M_x, M_y, M_z)$.

If condition 1 is fulfilled, and condition 2 is not, then a cell undergoes an *Internal cell event* of type S (the biological meaning of this cell event is described in Rule 5 below). If the condition 1 is not fulfilled, we determine the possibility to have cell event *Growth*. For that we check the condition:

If for at least one type of coding molecules $i$ ($i = 1, \ldots 8$) in a cell matrix, its deviation (in a row) satisfies $\sigma_r < e$, then there will be No cell event for this cell. Otherwise, the cell undergoes the cell event *Growth.*

We do not consider the cell event *Movement* in this simplified version of the program; this is a work in progress.

**Rule 2. Filling in elements in daughter cells during Cell event Division**

We will assume that one half of N sectors of each daughter cell (columns in a daughter cell matrix) will stay equal to the corresponding half of sectors of the mother cell (the ones related to outer sectors of a daughter cell), while each element in new (inner) sectors will be created *de novo* according to a universal rule, as schematically illustrated on Figure 3. This provides the partial inheritance of the information of the spectrum of a cell in the spectra of its daughter cells.

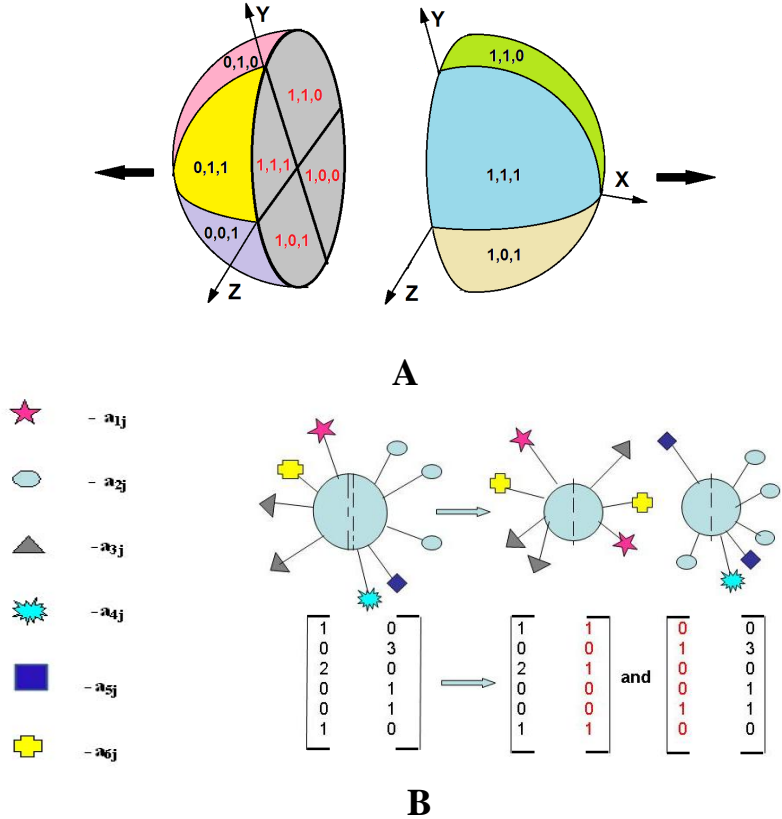

**Figure 3.** Illustration of a Rule for filling in the matrices, corresponding to daughter cells after division. **A**.- Spatial organization of the sectors on the two halves of mother cell before the division. **B**.-The simplified 2-sectors example of the Rule for filling in the matrices, corresponding to daughter cells after division. Each cell matrix containing 6 types of coding molecules on a cell surface with 2 sectors.

We suggest as a **Rule 2** that each new element in a matrix of a daughter cell is a function of the matrix of a mother cell. To discover this function as an operator, acting on a matrix (spectrum) of a mother cell and generating two new matrices of daughter cells, is a direction of our current theoretical work. For the time being, for a simplified version of a program, we suggest this function to be a linear combination of the same elements of type $i$ in all sectors of the mother's cell.

For the mathematical formalization of Rule 2, let us consider a coding matrix $a_{in}$ together with the spatial orientation of its $n$ columns (as sectors of a cell surface) in a zygotic/embryonic coordinate system (axes AP, DV, LR) in $R^3$ which we will replace by $xyz$ one. Then at any generation $t$ a cell spectrum will be described by the matrix:

$$\hat{a} = \left[ a^t_{i_{xyz}} \right],$$

where $a^t_{i_{xyz}}$ is given by the number of molecules of i-th type in a sector with spatial coordinates x, y, z.

Also, let us admit that having a division in L-direction (L = x, y, z), i.e., a cell division plane being perpendicular to the axis L, we will notate the daughter cells as $(L, \varepsilon), \varepsilon = 0,1$. This means, that the daughter cell, emerging in the positive direction of axis L gets index $\varepsilon = 1$, while the cell emerging in the opposite direction gets index $\varepsilon = 0$.

In this case **Rule 2** for filling in the elements of the spectrum of a $(L, \varepsilon)$ -daughter cell during Cell event *Division* will be written as:

$$a^{t+1}_{i_{xyz}} = a^t_{i_{xyz}} , \text{ if } x = \varepsilon,$$

(3)

$$a^{t+1}_{i_{xyz}} = c_1 a^t_{i_{xyz}} + c_2 a^t_{i_{\underline{x}yz}} + c_3 a^t_{i_{x\underline{y}z}} + c_4 a^t_{i_{\underline{xy}z}} + c_5 a^t_{i_{xy\underline{z}}} + c_6 a^t_{i_{\underline{x}y\underline{z}}} + c_7 a^t_{i_{x\underline{y}\underline{z}}} + c_8 a^t_{i_{\underline{xyz}}} , \text{ if } x = \underline{\varepsilon},$$

where: $\underline{x} = 1 - x; \underline{y} = 1 - y; \underline{z} = 1 - z; \underline{\varepsilon} = 1 - \varepsilon;$ $c_i$ are chosen parameters.

This means the conservation of half the information of a mother cell in half of the sectors of a daughter one, while in the sectors of another half of daughter cell each new element will be calculated as a linear combination of its amount in all sectors of the mother cell. The rule (3) is written for the axis x being the axis of a division plane. For all other cases before and after the application of this rule the program will perform the L - permutation, which is a permutation of sectors indexes which takes the index L to be the first, while the others move circularly.

Note that for each new element $a^{t+1}_{i_{xyz}}$ in a daughter cell, all odd coefficients $c_i$ in formula (3) correspond to the elements of the "mother" part of a coding matrix belonging to this cell, while the even coefficients $c_i$ - to the "mother" part of a coding matrix which will move to the other daughter cell after division.

Another important remark is that the law is written in such a manner that it provides for both L,0 - cell and L,1 - cell conservation of an outer half of information of a mother cell, while the part which will be filled in with new information will correspond to the novel sectors on a cell division septum (for plants) or on the stretched out parts of the cell membrane (for animals).

In general, to specify the Rule of filling in the elements (inheritance), the 8 parameters $c_1 \ldots c_8$ should be required. But we will consider biologically meaningful conditions/restrictions for $c_i$ in (3) which allow reduction of the number of chosen parameters.

**Rule 3. Complementarity of daughter matrices**

The **main hypothesis** about the conditions on filling in the elements is the requirement of *«Complementarity»* of nascent neighbor sectors in two daughter cells. According to our model, each of four sectors of a daughter cell which were filled in the course of division has a corresponding adjacent sector in another daughter cell, also filled during the same cell event Division. We will call each pair of these new-formed adjacent sectors the «twins».

The suggested Rule 3 of *«Complementarity»* proposes that the sum of each coding substance in two «twins» will be the same for all pairs of twins. This means that the sum

$$a_i^{t+1}{}_{1yz(L,0-cell)} + a_i^{t+1}{}_{0yz(L,1-cell)}$$

is independent of $y, z$. This is possible if:

$$c_1 + c_2 = c_3 + c_4 = c_5 + c_6 = c_7 + c_8 = C. \quad (4)$$

Let us refer to the parameter $C$ as the *parameter of Complementarity*. If we consider a total sum for a substance i in a cell:

$$A_i^t = \sum_{x,y,z} a_{i_{xyz}}^t, \quad (5)$$

then from (3), (4) and (5) we will get:

$$a_i^{t+1}{}_{1yz(L,0-cell)} + a_i^{t+1}{}_{0yz(L,1-cell)} = C A_i^t \quad (6)$$

for any y, z and any substance i.

Furthermore, we can assume the existence of additional restrictions on the coefficients in formula (3) depending on the process of twin sector formation. For example, we can consider symmetry or spirality pattern of twin sector formation. This will give the following relations for the coefficients:

- in the case of the Symmetrical process:

$$c_3 = c_5; \quad c_4 = c_6 \tag{7}$$

- in the case of the *Spiral process:*

$$c_4 = c_5 = 0 \quad \text{for the left one} \tag{8}$$

$$c_3 = c_6 = 0 \quad \text{for the right one.} \tag{9}$$

**Introducing the generalized parameters λ and μ.**

The parameter of complementarity C from formula (4) may be written as:

$$C = (c_1 + c_2 + c_3 + c_4 + c_5 + c_6 + c_7 + c_8)/4 = \frac{2\lambda - 1}{4}, \tag{10}$$

where $\lambda$ is parameter of inheritance, calculated from coefficients $c_i$, and characterizing an average change of a sum of each element in two daughter cells after division.

Indeed, from (3) we obtain for each coding molecule (substance) i:

$$A_i^{t+1}{}_{(L,0-cell)} + A_i^{t+1}{}_{(L,1-cell)} = 2\lambda A_i^t. \tag{11}$$

We will posit that $\lambda < 1$, which means that the total sum of all elements in a spectrum of a cell decreases with generations. This comes from the fact that the potency of cells actually decreases in a course of divisions and differentiation and thus the amount of "status-significant" elements (substances) on its surface should decrease too.

For introduction of the parameter μ, we will re-write a moment $M$ of a cell in a direction L, $(L = x, y, z)$ from formula (1) as:

$$M_L = \sum_i M_{i \cdot L}, \tag{12}$$

where $M_{i \cdot L}$ is a Moment of a substance $i$ in each direction:

$$\begin{aligned} M_{i \cdot x} &= \sum_{y,z}(a_{i1yz} - a_{i0yz}) \\ M_{i \cdot y} &= \sum_{x,z}(a_{ix1z} - a_{ix0z}). \\ M_{i \cdot z} &= \sum_{x,y}(a_{ixy1} - a_{ixy0}) \end{aligned}$$

Then taking into consideration formula (3), we see that for each substance $i$ will hold:

$$M_i^{t+1}{}_{L(L,0-cell)} + M_i^{t+1}{}_{L(L,1-cell)} = 2\mu M_{i\,L}^t, \tag{13}$$

where µ is a coefficient which gives an average of the change of longitudinal moment in two daughter cells after division with L the direction of cell division. The parameter µ influences the dispersion of daughter cells by the sum of their elements and thus their possible diverse differentiation. Also, it can be seen that for µ > 1, the average longitudinal moment of the daughter cells increases thus increasing the chance of continuation of the division in the same direction as in the mother cell.

We can see that the parameters λ and µ are related with the coefficients $c_i$:

$$c_1 + c_3 + c_5 + c_7 = \lambda + \mu - 1, \tag{14}$$

$$c_2 + c_4 + c_6 + c_8 = \lambda - \mu, \tag{15}$$

which means a possibility to decrease the number of required parameters of the model. The specification of complementarity (4) according to formulas (14), (15) decreases the additional (to λ and µ) parameters to 3 (for example, the coefficients $c_1$, $c_3$, $c_5$). The condition of symmetry of inheritance requires setting only 2 additional coefficients ($c_1$ and $c_3$), while the condition of spirality– only the single coefficient $c_1$.

Next we can obtain the general formulas for the total sum of each coding substance in daughter cells:

$$\begin{aligned} A_{i\,(L,0-cell)}^{t+1} &= \lambda A_i^t - (1-\mu)M_{i\,L}^t \\ A_{i\,(L,1-cell)}^{t+1} &= \lambda A_i^t + (1-\mu)M_{i\,L}^t \end{aligned}, \tag{16}$$

which is important for detecting cell differentiation; and for the longitudinal moment:

$$\begin{aligned} M_{i\,(L,0-cell)}^{t+1} &= -(1-\lambda)A_i^t + \mu M_{i\,L}^t \\ M_{i\,(L,1-cell)}^{t+1} &= (1-\lambda)A_i^t + \mu M_{i\,L}^t \end{aligned}, \tag{17}$$

which is important for determining the further direction of division of daughter cells. Also, both formulas (16) and (17) are important for calculation concerning signaling (see section 3.3s).

**Rule 4. Changes of cell surface markers during cell event Growth.**

For the rule of the changes of surface markers during cell event Growth between its birth and disappearance (when it divides), we consider two halves of the surface in the $M_{max}$ direction.

In each column of the half for which the sum of all components (A,G,F,K,P,D,X,Y) is smaller, we apply the formula, decreasing the minimal element in a column:

$$a_{min}^{xyz} - e,$$

where $e$ is calculated according to (2).

In each column of the half for which the sum of all components (A,G,F,K,P,D,X,Y) is bigger, we apply the formula, increasing the maximal element in a column:

$$a_{max}^{xyz} + e.$$

The biological basis of this rule is that in most cases during a period of growth, a cell prepares itself for undergoing cell division, thus, the proposed changes of cell surface markers during cell event "Growth" allow the cell to increase the heterogeneity of its markers and thus to increase (according to the condition 2 (2)) its ability to satisfy the conditions for the choice of the cell event "Division". However, as it is also frequently occurs in many cases that no cell division will happen to a cell after its growth, if its heterogeneity increases not significantly enough to satisfy condition (2), or if the other condition (1) is not fulfilled.

**Rule 5. Internal cell event of type S**

During Internal cell event of type S a code of a cell is changed without any external cell event for the cell. Internal cell events are regarded as the necessary steps in realization of the determined developmental program, and they are included in the model in order to reflect the response of a cell to a set of biochemical factors, which **should come to a cell** at this step in a case of normal development (and as a result, the matrix will be changed). But it is assumed that Internal cell events can also occur as a response to the abnormal environmental changes. One example of it is presented in the Rule 10 (see below), when the cell changes its matrix as a response to the changed information from its neighbor cells. The change of the code during an Internal cell event of type S is suggested as follows:

We consider Left and Right halves of the surface in the $M_{max}$ direction.

1. If the standard deviation $\sigma$ of all elements in any one half of the surface in the $M_{max}$ direction satisfies the condition

$$\sigma > e,$$

while the other does not, then it will be the changes in that half which do NOT satisfy the condition, according to the formula:

$$a_{max} + e$$
$$a_{min} - e,$$

where $e$ is calculated according to (2).

2. If BOTH deviations $\sigma_{left}$ and $\sigma_{right}$ of all elements in the Left and Right halves of the surface in the $M_{max}$ direction satisfy

$$\sigma \leq e,$$

then in each half the changes will be:

$$a_{max} - e,$$
$$a_{min} + e.$$

The biological basis of this rule is that during an Internal cell event of type S the change of cell matrix renders a cell capable of performing a new cell event which was not possible with the former matrix. The rule proposes that if the heterogeneity of the matrix is considerably large, then a cell has the potential for a chain of upcoming cell events; as a consequence, during an Internal cell event its heterogeneity will be continuously increasing, thus allowing realization (according to condition 2) of next cell event (preferentially, a cell division, but not necessarily). Alternatively, if the heterogeneity of the whole matrix is already rather small, then in the course of an Internal cell event a cell will acquire (by further decreasing its heterogeneity) the conditions for "no cell event" status meaning a cell silencing.

**Rule 6. Determination of apoptosis.**

If all elements â in at least 1 sector of a matrix satisfy the inequality:

$$\left|a^{t}_{i\,xyz}\right| < e,$$

then a cell will undergo apoptosis. The cells which undergo apoptosis are marked in black color and remain unchanged during all further development of the embryo, thus reflecting the possible emerging cavities in the body of an embryo, which are the most usual result of apoptosis. The underlying biological sense of this rule is that the parameter $e$ can be considered as the unit of meaningful value of an element in a matrix (a meaningful amount of a particular type of coding molecules in a sector). Thus, if at least in 1 sector of a matrix the amount of all coding molecules decreases below this level, it can be considered as a mark for cell death (demolition).

**Rule 7. Determination of differentiation.**

We will consider the process of differentiation as an *Internal cell event* of type D, which does not acquire a unit time period by itself, and thus can coincide with all other types of cell events. The differentiation status of a cell depends on the content of its matrix and is detected by the program at the moment of cell appearance. If, according to the suggested rule, a cell undergoes differentiation, then this cell will be marked by one of 8 different colors, reflecting its differentiation status (presented on Figure 2).

The Rules for differentiation are determined by a proportion $d_i$ of a substance $i$ in a total sum of all substances, $0 < d_i < 1$.

To determine if a cell undergoes a cell event *Differentiation*, the program performs the following check for each cell at the end of each cell event:

1. Calculate a sum of all coding elements in the matrix of a cell $A^t = \sum_i A_i^t$, and calculate a proportion of each substance i in $A^t$ cell: $d_i = |A_i^t/A^t|$.
2. If $d_i$ is the maximal proportion among all substances $k$:

$$d_i = \left|\frac{A_i^t}{A^t}\right| = max_k \left|\frac{A_k^t}{A^t}\right| \text{ and } d_i > d,$$

   where $d$ is a chosen parameter, then a cell will get differentiation status of i-type.
3. When cell gets a differentiated status, a novel mode of inheritance is switched on for this cell. This means that for i-differentiated cells (cells with differentiation status of i-type), the parameter $\lambda$ will be replaced by $\lambda = 1$ in all formulas for calculating the parameters of rules.

Notice that the choice $\mu = 1$ provides the same differentiation status of the daughter cells as that of the mother independently of $\lambda$ (see (16)).

**Rule 8. Cell adhesion.**

Adhesion is described as the force $F$ between two intercommunicating cells, which can have three possible states: strong, medium and zero, depending on the content of coding matrixes of the two contacting cells. The medium adhesion occurs when cells can change their mutual position, but do not come off from each other, the strong one - when cells cannot change their relative position. The state zero corresponds to the case when a particular cell has no adhesion to other cells, thus enabling this cell to move in the body passing by other cells (which is the case, for example, for stem cells in Planaria). For such "zero adhesion" cases, the adhesion depends only on the content of the matrix of one cell having "zero adhesion", independently of the matrix of its neighbors. The current program considers only medium adhesion for all cells. Other cases are work in progress.

### 3.3 Signaling

According to our theoretical assumptions [1], the signaling between cells depends on epigenetic codes (matrices) of neighbor cells. Namely, we call *signaling* the transmission by each cell its own epigenetic spectrum to a collection of its neighbors. In the model, the response to signaling includes a set of simple but biologically relevant rules. Briefly, at each time-step, the cell detects its normal or abnormal position relative to the signal received from neighbor cells, based on the information of their epigenetic matrices. The main criteria of this detection are the complementarity of matrices of neighbor cells and their relative positions. In the case of detection of normal signal, the programmed development of a cell occurs according to the set of Rules 1-8. In the case of detection of abnormal signal (e.g., in the cases of amputation, transplantation, malfunction and other "crises") a process of coordinated change of cell fate such as cell dedifferentiation with further regeneration according to the suggested rules (**Rules 9,10**) will be activated.

For formulating the rules for detecting signaling between cells, we should make 2 remarks:

(1) We assume that in the process of normal development without programmed cell event *Movement* the distance between daughter cells is assumed to be rather small, i.e., not exceeding 3R, where R is a cell radius.

(2) As was mentioned before, we call each pair of new-formed adjacent sectors of two daughter cells the «twins»; thus, each cell division produces 4 pairs of «twins», and each cell in an embryo can have maximum number of 8 «twins» relations with its neighbors.

Using these statements, we suggest the following rules (**Rules 9) for detecting signaling between cells:**

1.  If all distances T between «twins» sectors of a cell and its «twins» neighbors do not exceed R, then a cell undergoes normal development and receives **no abnormal** signal.

2.  If at least one distance T between «twins» sectors of a cell and its neighbors are greater than R but do not exceed 3R, then a cell receives a **small abnormal** signal.

3.  If in a cell which has N «twins» relations and the distance T between L «twins» sectors exceeds 3R, where $0 < L \leq N/2$, then a cell receives a **middle abnormal** signal.

4. If in a cell which has N «twins» relations, the distance T between L «twins» sectors exceeds 3R, where L > $N/2$, then a cell receives a **strong abnormal** signal.

We propose the following rules **(Rules 10) for the response to the abnormal ("crisis") signaling:**

In the case of a **small signal,** the cell event will *depend on* the matrix â of a cell, as is supposed to be for the optimal (coded) development.

In the case of a **middle signal,** the cell event will be *independent of* the matrix of a cell, and can be one of the 3 scenarios:

-cell movement,

-de-differentiation (proliferation),

-stagnation (no cell event).

For the time being, the choice of the exact scenario should be chosen by the experimentalist (see description in Supplemental materials), and will be the same for all cells receiving a middle signal.

In order to provide a response to a **strong signal**, the program calculates for a cell a "*determining matrix*" $â_d$, which is a "complementary to adjacent" matrix for the matrix $â$, i.e., $â_d = (â_a)^{compl}$.

We will determine an "*adjacent matrix*" $â_a$ for a cell having a matrix $â$ as a matrix with 8 columns corresponding to 8 adjacent sectors of neighboring cells (for simplification we consider that 1 sector of a cell is adjacent to 1 sector of one of neighboring cells).

A *complementary* matrix to any given matrix is constructed from eight complementary sectors to eight sectors of the matrix of a cell. According to (4), (6), (13), (16), and (17), the formula for the calculation of a complementary sector to a given sector of a cell will be:

$$(a^t_{i\ xyz})^{compl} = \mu \frac{C}{S_0} A^t_i - a^t_{i\ xyz}, \qquad (18)$$

where $S_0$ is a sum of odd coefficients $c_i$ (see (13)). Actually, this is a simplified isotropic version of the formula, which derives from the calculations and includes a dependence of the complementary sector on the moment $M^t_{i\ L}$ of the preceding division:

$$(a^t_{i\ xyz})^{compl} = \mu \frac{C}{S_0} A^t_i \pm \frac{1-\mu}{S_0} M^t_{i\ L} - a^t_{i\ xyz}, \qquad (19)$$

where the index L=0,1 indicates different daughter cells, "+" corresponds to 0, "-" to 1.

Indeed, from (4), (6) we have:

$$(a_i^{t+1}{}_{xyz})^{compl} + a_i^{t+1}{}_{xyz} = CA_i^t, \qquad (20)$$

and the transition from $A_i^t$ to $A_i^{t+1}$ according to (13), (16), (17) results in two formulas for different the daughter cells:

$$A_i^t = \frac{\mu}{S_0} A_i^{t+1}{}_{L(L,0-cell)} + \frac{1-\mu}{S_0} M_i^{t+1}{}_{L(L,0\ cell)}, \qquad (21a)$$

$$A_i^t = \frac{\mu}{S_0} A_i^{t+1}{}_{L(L,1-cell)} - \frac{1-\mu}{S_0} M_i^{t+1}{}_{L(L,1\ cell)}. \qquad (21b)$$

Taking $A_i^t$ as an average of (21a) and (21b), we get (18), allowing direct calculation of a complementary sector to a given one without additional (moment) information.

Next using (18) (or, if possible, (21a), (21a)), the eight sectors of a matrix $\hat{a}_d$ are calculated, which provides a "*complementary to adjacent*" matrix for a matrix $\hat{a}$.

Using these calculations, we formulate **the rule for the response to a strong abnormal signal:**

If the *total sum* A of all elements in a matrix $\hat{a}$ of a cell and the *total sum* $A_d$ of all elements in a matrix $\hat{a}_d$ satisfy an inequality $A < A_d$, a cell will *undergo cell death* (apoptosis). If $A > A_d$, the cell A will *convert its matrix from* $\hat{a}$ *to* $\hat{a}_d$. This rule means that in the case of strong signal (strong difference between its own spectrum and the "adjacent spectrum" of neighbors) the cell "older" than its environment is dying, while the behavior of the "younger" cell is totally determined by this environment, each one complementary to one sector of the matrix $\hat{a}_a$, $\hat{a}_d = (\hat{a}_a)^{compl}$.

The next cell event in this situation is performed according to the programmed development, i.e., the set of Rules 1-8) with the newly acquired matrix $\hat{a}_d$.

### 4. Results and Discussion

#### 4.1. Computational experiments-description and general results

The computational experiments using the Morphogenesis Software starts with the generation of an initial cell of an organism (zygote) with an assigned matrix. Next the development of an embryo from zygote is modeled following the set of suggested rules 1-8 up to the formation of early stages of embryogenesis (up to around 1500 cells). The influence of variations of

(1) different sets of parameters λ, μ, C, $m_h$, P, X, d;

(2) the possible choices of the type of the initial matrix (random one or diagonal random one with different coefficient) ;

(3) possible choices of the mode of complementarity (symmetrical or spiral);

(4) the modes of signaling on the developing computational embryo were studied by changing them one by one with the same zygotic matrix.

This study was repeated for a few hundred zygotic matrices, which gives the possibility to determine the best values of parameters (1) and the best choices of program modes (2), (3), (4) which display the evolution of computational embryos in best possible accordance with normal development (e.g., excluding the interruption of the development, appearance of definitely abnormal structures, unrealistically immense amount of apoptosis, thus demonstrating realistic stages for starting the processes of differentiation and structure formation). Next, having these parameters and program modes fixed, we studied the effect of the variation of zygotic matrices on the appearing shape of an embryo.

The computational experiments demonstrated the following results.

First, the computational experiments using the Morphogenesis Software have shown that any change of one element in a zygotic matrix keeping all other parameters the same causes quite visible (and sometimes very essential) changes in the resulting shape of a computational embryo. This means the existence of a one-to-one dependence of pattern formation (a shape of a developing embryo and differentiation of its cells) on a spectrum of coding molecules on the initial (zygotic) cell matrix.

Second, we have found that any developmental tree with pre-determined parameters of cell events corresponds to specific embryo morphology at each time slice.

Third, we have found that among millions of numerical embryos, starting from initial cells (zygotes) with randomly generated coding matrices and next undergoing development following the set of rules 1-10 with different sets of parameters $\lambda, \mu, C, m_h, P, X, d$, there exist several numerical embryos with the shapes well approximating the shapes of actual embryos at approximately equivalent stages of development (around 1000-1500 cells). The good approximation of the shapes of these embryos, belonging to different plant and animal taxa, means also a similarity in differentiation status of various tissues and structures.

In order to prove this similarity by precise mathematical methodology, and also to illustrate the second statement, we have performed a comparison of the developmental trees of the actual organisms with the computational trees produced by the Morphogenesis Software, followed by the comparison of corresponding shapes of the embryos.

### 4.2. Development of actual organisms, presented as a tree of cell events

For the purpose of model validation, we have studied the development of early stages of embryogenesis of actual organisms, with its next formalization as a graph (i.e., a rooted graph or tree) with vertices corresponding to cells in particular cell states, and edges corresponding to cell events, representing cell fates. Namely, the anatomical sections of a developing embryos of three angiosperm species belonging to different classes and families: *Miryophyllum specatum* L. (Haloragaceae, dicot) [31], *Polygala major* Jacq. (Polygalaceae, dicot) and *Triglochin palustre* L. (Juncaginaceae, monocot) [32], were investigated. The analysis of these anatomical sections enables revealing cell events reflecting the fate of each cell and results in the construction of corresponding developmental graphs (trees) for these species, together with the reconstructions of a 3-dimensional shape of the embryos at progressive developmental stages (Fig. 4,5,6).

Each horizontal layer of a graph corresponds to a **level** of cell development, where each new level indicates a step (time period) during which at least one cell event has occurred at least in one cell. During the observed period of development only three types of cell events were observed: cell growth, cell division and cell differentiation, while no cell movement and no cell death events were noted.

On the graphs, a cell event *Growth* is manifest as an edge with a number coefficient showing cell enlargement in size. The planes of cell division, which are determined according to coordinate system established for the zygote, are indicated by the labels on the two edges emerging from a corresponding vertex: "x" corresponds to the transverse plane of cell division, "y" corresponds to the longitude plane of cell division, and "z" designates a division in a plane perpendicular to the xy plane. The spatial distinction in a pair of daughter cells is reflected on the graphs in the following way: for the axis x, the left and right vertices of the graph (emerging from the same vertex) directly correspond to the left and right positions of daughter cells related to axis x; for the axis y, the left vertex corresponds to the upper cell, while the right vertex corresponds to the lower cell; the vertices with circles inside indicate those cells from pairs of daughter cells divided in the z-direction which correspond to the bottom part of z-axis.

The edge presented as a dotted line indicates the situation when during a particular step no cell event has happened for a cell.

The first event of differentiation in an embryo is differentiation of an embryoderm, an outer layer of an embryo. In the developmental graph built for *Polygala major*, the vertices which correspond to the differentiated cells of embryoderm are colored in black.

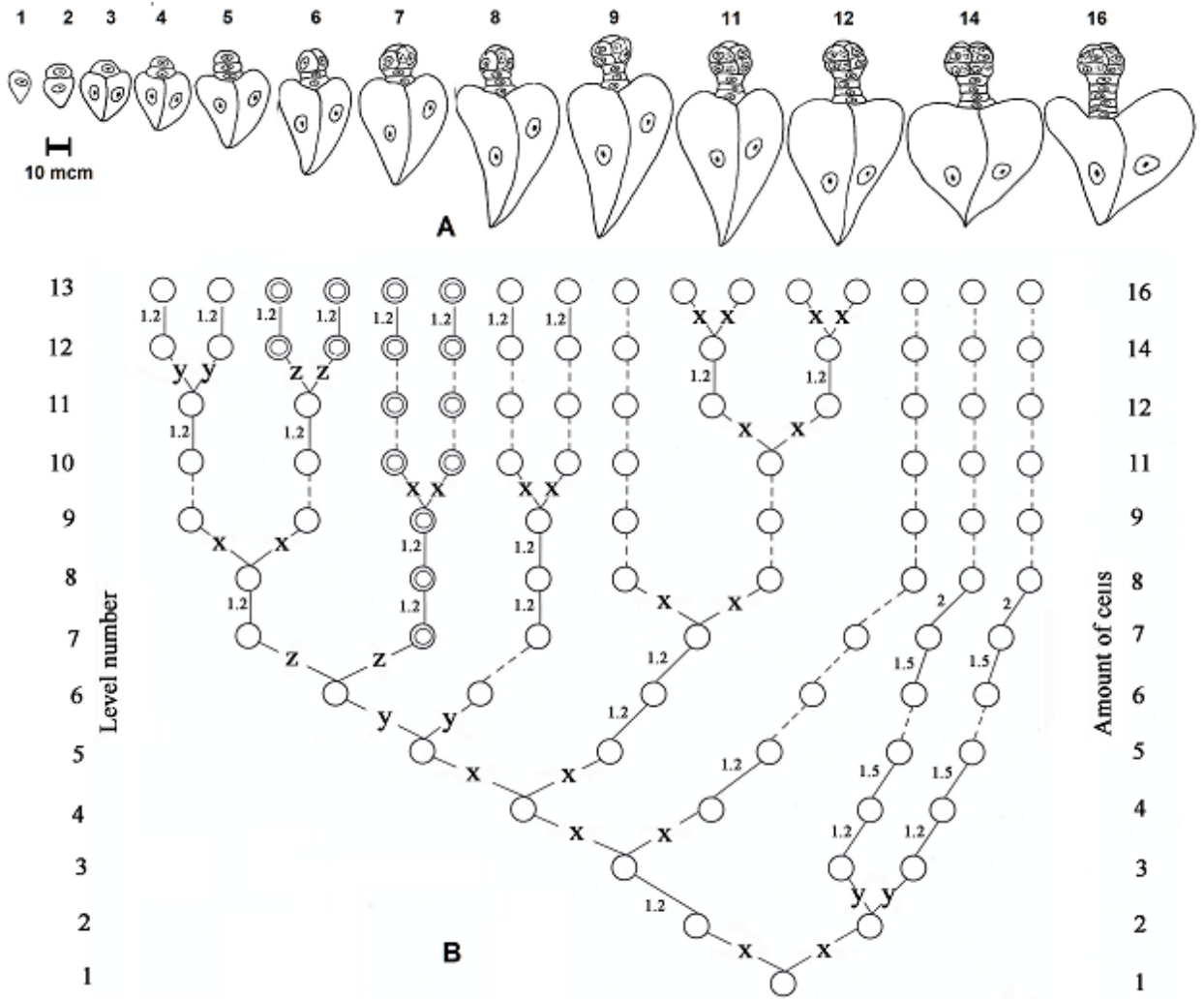

**Figure 4**. Embryogenesis of Myriophyllym specatum. **A**. The scheme of the first stages up to the 16-cellular proembryo. **B**. The corresponding graph (tree) of cell events. The description of the labels on the graph is in the text.

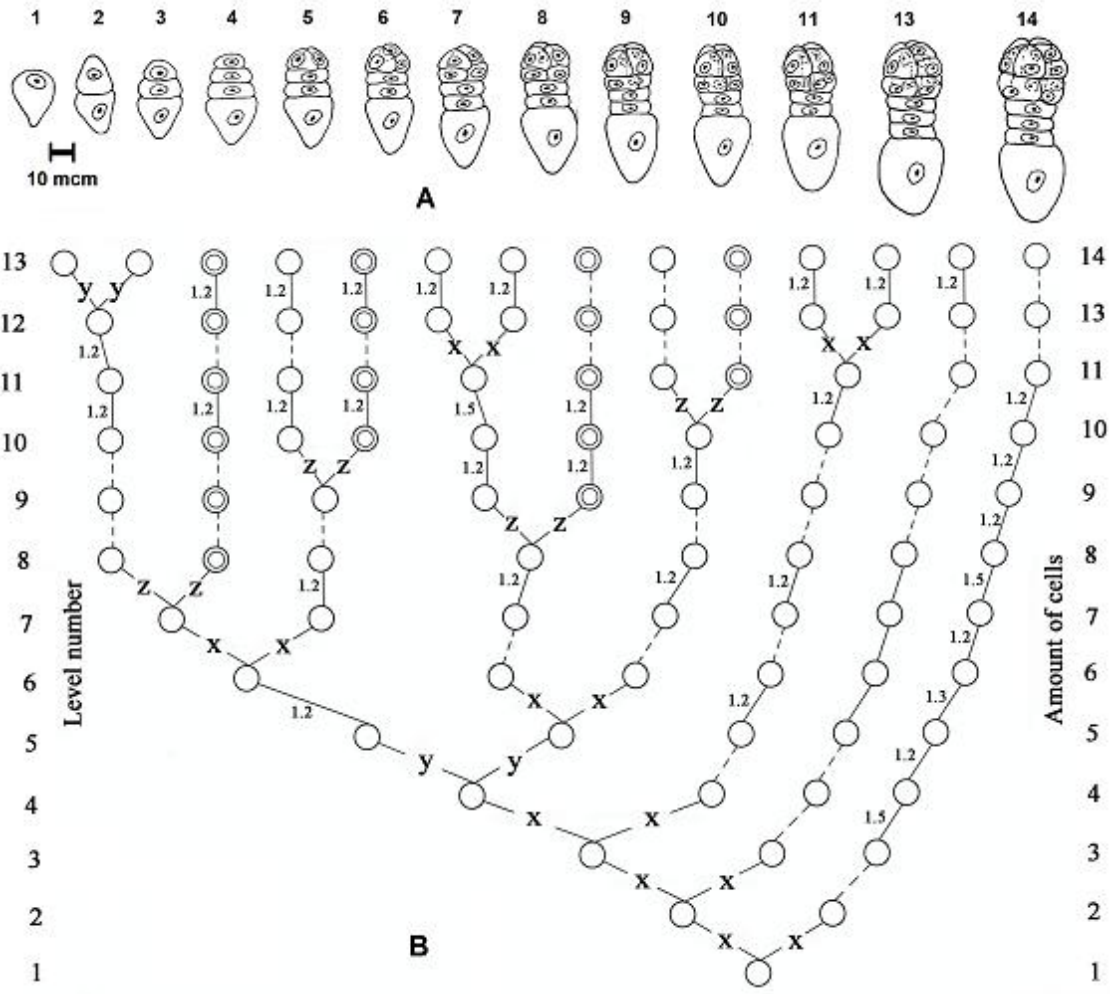

**Figure 5.** Embryogenesis of Triglochin palustre. **A**. The scheme of the first stages up to the 14-cellular proembryo. **B**. The corresponding graph (tree) of cell events. The description of the labels on the graph is in the text.

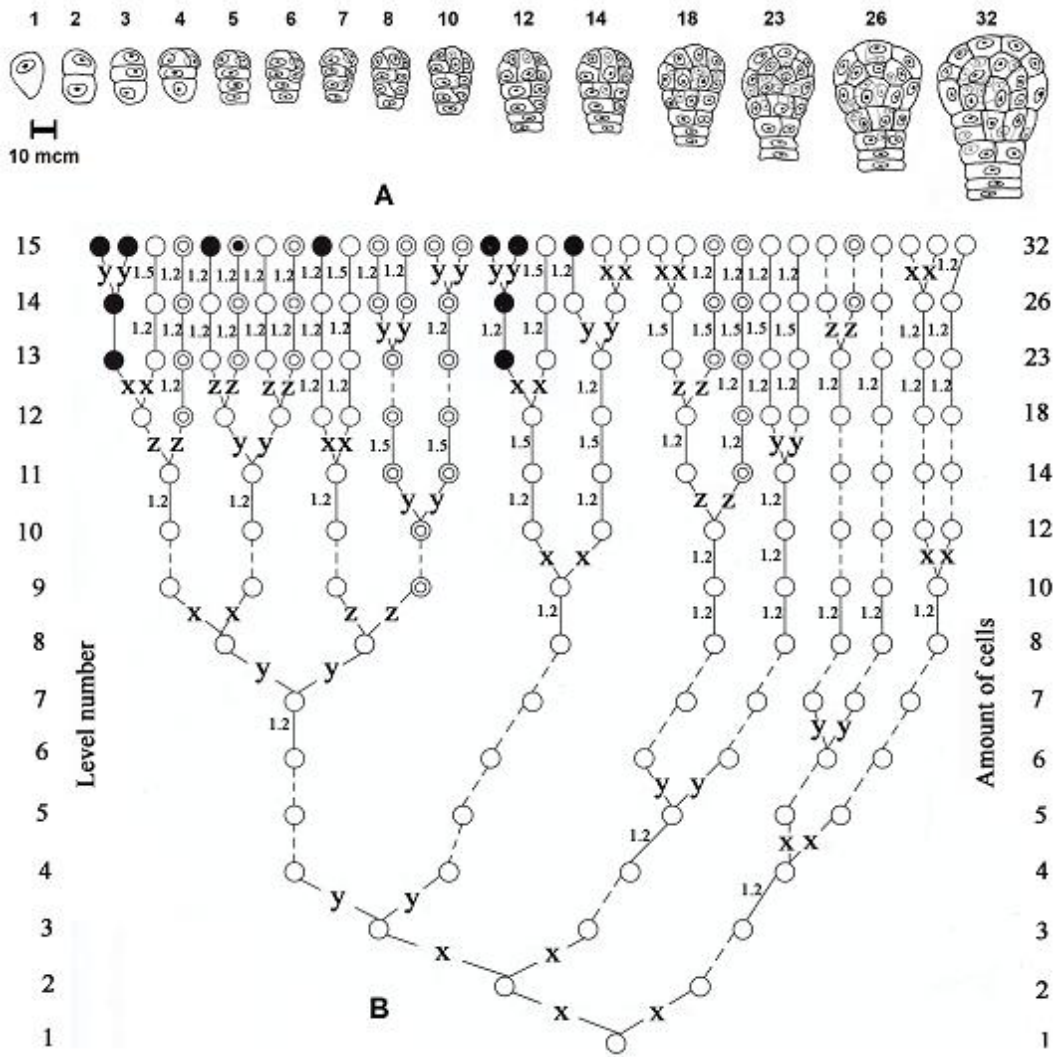

**Figure 6.** Embryogenesis of Polygala major. **A**. The scheme of the first stages up to proembryo. **B**. The corresponding graph (tree) of cell events. The description of the labels on the graph is given in the main text.

### 4.3. Comparison of actual and computational trees and shapes

The trees, obtained from tens of millions of randomly generated zygotic matrices, developing into the embryos by the Morphogenesis Software, were compared with each of the three developmental trees of the actual organisms, presented in Figs 4,5,6. The source code for matrix generation, embryo growth and trees comparison can be found at:

https://github.com/nickbessonov/morphogenesis-article/blob/master/Sim16_01.rar

and the algorithm for the developmental graphs (trees) comparison is presented in Supplemental Materials. The embryos were developed with the "Random" choice of the initial matrix, the constant set of all parameters ($\lambda=0,965$; $\mu=0,655$; $C=\frac{(2\lambda-1)}{4}$; $c_1=c_3=\frac{2}{3}C$; $m_h=100$; $P=0,75$; $X=0,1$; $d=0,2$), the symmetry pattern of matrices complementarity and with the request to follow the rules for normal signaling. The best zygotic matrices found for the developmental trees, closest to the actual ones, and the corresponding trees, are presented in Supplemental Materials. As the computational trees includes edges for both "Internal cell event" and "No cell event", while the information about the actual trees provides "No cell event" edges only (as currently we do not have information about molecular markers, providing matrices, and thus, about their change, i.e., "Internal cell event"), for the tree comparison algorithm, we have considered both types of events to be equivalent.

It is important to note that these computations were based on having the simplest case of initial matrix (8X8) which gives the number of possible matrices as $101^{64}$, which provides much less than 0.1% of the maximal amount of possible trees (around $5^{(2^{n+1}-1)}$, where 5 is the number of possible cell events in the actual cases of plant species (cell division in three directions (x,y,z), growth and no cell event), n is the number of levels of the tree). And even for this very simplified case (only 8 possible coding molecules, while in the theory we assume all possible combinations of mono- ,di- and three-saccharides, and only 8 sectors of the cell surface, which should be increased) we have achieved the putative zygotic matrices, providing the trees with 0,89% of similarity with the actual ones. This provides a good proof-of-concept and the possibility to search next for a more complex matrices and allow a complete correspondence of computational trees to actual ones.

Next we have checked that the computational trees, identical to the trees of three presented plant species, give the same embryo shapes as the actual species. The results produced by the Morphogenesis Software up to levels 13-15 are shown in Figures 7,8,9. The labels of the edges on these graphs (trees) are the same as for the developmental trees presented for the actual embryos on the Figures 4,5,6, except for the "No cell event" edges marked with "0" (instead of dashed line on the trees for actual embryos) and "Internal cell event" edges marked with "1" (this step exists on computational trees only and, as mentioned above, is considered to be equivalent to "No cell event" for the comparison algorithms).

The obtained shapes of numerical embryos (Figures 7,8,9) were compared to the corresponding embryo shapes of three plant species, presented on the Figs 4,5,6 using standard image comparison software (https://opencv.org). The results of comparison have shown high degree of similarity for all three plant species.

Namely, the distance between the original shapes and the computational ones (calculated as a simple Hausdorff distance measure between shapes defined by contours) were scored as 160, 54, 34 units correspondingly, while a threshold of good similarity of two images stated in the software is 200 units and the best matching is stated to be below 50 units.

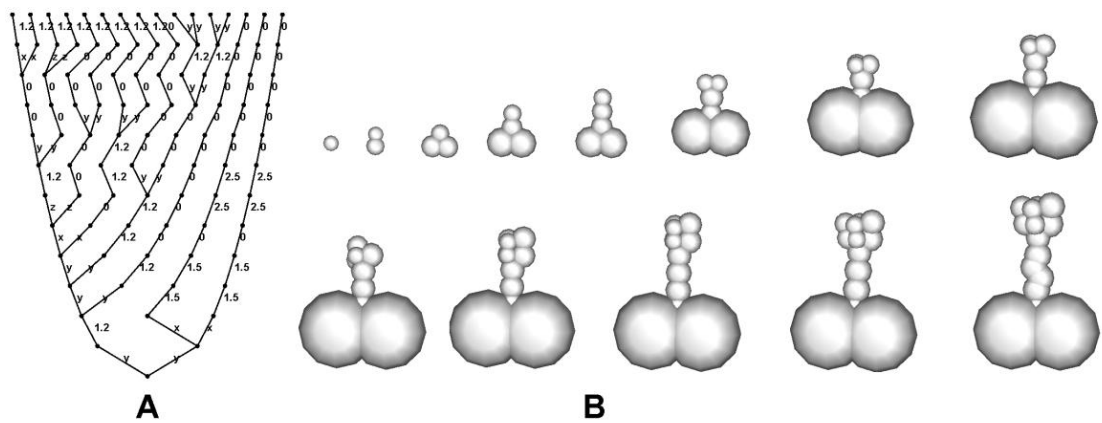

**Figure 7**. Embryogenesis of Myriophyllym specatum produced by the Morphogenesis Software. A. Computational tree, similar to the actual tree presented on the Figure 4. B. Corresponding computational embryos up to the level 13. The description of the labels on the graph is in the text.

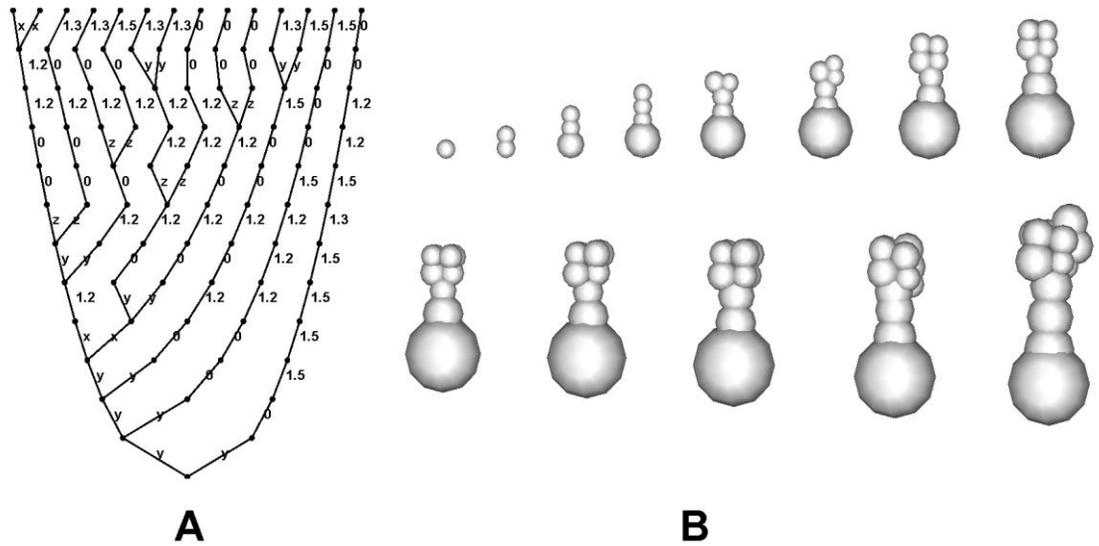

**Figure 8**. Embryogenesis of Triglochin palustre produced by the Morphogenesis Software. A. Computational tree, similar to the actual tree presented on the Figure 6. B. Corresponding computational embryos up to the level 13. The description of the labels on the graph is in the text.

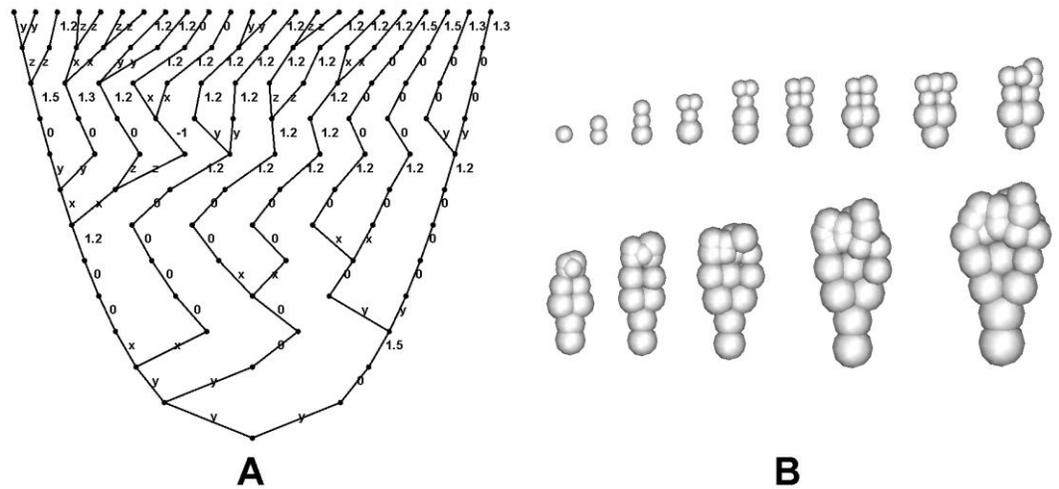

**Figure 9**. Embryogenesis of Polygala major produced by the Morphogenesis Software. A. Computational tree, similar to the actual tree presented on the Figure 6. B. Corresponding computational embryos up to the level 15. The description of the labels on the graph is in the text.

This approach gives precise numerical results of comparison of the developmental processes, leading to the similar shapes of the objects, thus confirming the suggested theory and the acceptability of the chosen simplified rules.

**Discussion and Conclusions**

In this work we provide a frame of mathematical formalism and a corresponding software demonstrating a connection between a presumed epigenetic code, a consequent cell event for a cell and the geometry of an embryo. As an apparent good result, by application of the software to millions of randomly generated zygotic matrices representing such a code, we were able to find several numerical embryos with shapes well approximating the shapes of different actual embryos (from plant and animal kingdoms) containing up to 1000-1500 cells. However, the complexity of these embryos and an ambiguity of the data taken from published sources do not allow performing the precise comparison analysis required for validation of the model. Therefore, we have validated the conceptual idea underlying the model on three plant organisms, whose early developmental stages were examined in a detailed manner according to the suggested formalization, thus enabling the precise numerical comparison.

It is important to note that the main purposes of the Morphogenesis Software is to test the hypothesis of existence of epigenetic code controlling the geometrical shape of an embryo and to provide a tool for analyzing its compulsory characteristics and possible constraints of the interconnection of such a code and embryo morphology. As was already stated, we would like to propose the idea of this code in much wider context than our particular suggestion about (1) cell surface location of it and (2) the glycoconjugate residue nature of it. Thus, the software provides a modeling framework for testing manifold different conjectures of the nature of the epigenetic code formalized in the manner of suggested matrix, and the possibility to check its influence on the emerging shape. These different conjectures on the nature of the epigenetic code may include DNA and chromatin modifications (most commonly associated with epigenetic control ), electrical signals, mechanical tensions, other assumptions, or their meaningful combination.

Next we plan to apply our software to investigate development in other organisms. One issue is to check the model on C. elegans, whose early development is very well established and analyzed, and based on this, to reveal some compulsory characteristics of the proposed epigenetic code. The other prospective application is to elucidate, by using the software, the key principals of the process of regeneration in Planaria.

Apparently, for both of these perspectives the important consideration is the molecular implementation of the proposed epigenetic code, and its proper formalization in the suggested form of a matrix. Biological experiments might test the particular conjecture about the existence of a cell surface epigenetic code, and if so, provide concrete data for building a mathematical model based on a set of concrete data; this is our current work in progress.

The software comprises a concept of "crisis signaling" represented as a linear combination of epigenetic codes of neighbor cells. We assume that this signaling can account for changes in cell fate as within the normal course of development as well as in crises like regeneration, transplantation, cell isolation, etc. For the time being, we do not aim to point out the exact molecular mechanisms involved in implementation of this signaling and a fortiori do not aim to model all sets of different types of signals and signaling mechanisms in living organisms and their interconnection with this "crisis signaling".

Currently we are moving further ahead from the simplified version of the rules and developing an advanced mathematical approach suitable to decipher the actual rules (*developmental laws)* for converting the coded morphogenetic information into instructive signals for cell events.

**Materials and Methods**

The Morphogenesis Software is constructed for exploring problems related to the processes of morphogenesis. The program consists of the executable file Morphogenesis.exe and is designed to work under the operating system Windows XP and above.

The development of an embryo from a zygote is modeled in a three-dimensional computational region, the size of which increases automatically during the course of embryo development. Cells are modeled as spheres, having a set of constant and variable parameters.

The computational time (CPU time) for a growth of organism consisting from several thousand cells is equal approximately to several seconds.

For each case of modeling of the development from the zygote one should specify in a program:
- parameters $\lambda$, $\mu$, $c_1$, $c_3$ (for symmetrical case), $m_h$, P, X, d;

- the possible choices of the type of the initial matrix (random one or diagonal random one with different coefficient);

- possible choices of the mode of complementarity (symmetrical or spiral);

- the modes of signaling.

The instruction for using the software and managing the parameters and the modes of the model is given in Supplemental Materials.

**Main parameters of the model visualization**

We specify a set of parameters $k_f, h_1, h_2$ that determine the spatial cell layout. The coefficient $k_f$ reflects the force acting between cells, while $h_1, h_2$ are parameters ($h_1 < h_2$), imposing the conditions on the mutual arrangement of cells. The set of $h_1, h_2$ is detected for each pair of neighbor cells at each time step and considered to be $h_1 = R_1 + R_2$, $h_2 = 1{,}5 h_1$, where $R_1$ and $R_2$ are the radii of these neighbor cells.

At each time step, we calculate the distance $h$ between neighbor cells 1 and 2, and the forces acting between them: $\vec{F_1} = \vec{e}f$, $\vec{F_2} = -\vec{e}f$, where

$$f = \begin{cases} k_f \left(1 - \frac{h_1}{h}\right)\left(1 - \frac{h_2}{h}\right), \text{at } 0 < h \leq h_2, (h_1 < h_2) \\ 0, \text{at } h > h_2 \end{cases}, \qquad (22)$$

where $h = |\vec{r_2} - \vec{r_1}|$, $\vec{e} = (\vec{r_2} - \vec{r_1})/|\vec{r_2} - \vec{r_1}|$, $\vec{r_1}, \vec{r_2}$ are radial vectors of cells 1 and 2 in a global coordinate system.

The conditions (22) means that when cells become unreasonably close to make the inequality $h < h_1$ be valid, then the force f begins to repel the cells. Conversely, if the inequality $h_1 < h < h_2$ becomes true, then the force f begins to pull the cells together. In this way the automatic regulation of the positioning of neighboring cells at a given equilibrium distance $h \approx h_1$ is provided. If the distance between cells becomes too big (larger than $h_2$), then the force f is assumed to be zero, thus excluding the interaction between these cells.

Accordingly, at each time step after the forces between all cells are determined, the corrective movement of each cell is carried out according to the equation

$$m\frac{d\vec{v}}{dt} = -p\vec{v} + \vec{F}, \qquad \vec{v} = \frac{d\vec{r}}{dt}, \qquad (23)$$

where $\vec{F}$ is a resulting force, m is a mass and $\vec{v}$ is a velocity of a cell, $p$ is the viscosity of the medium (which may be non-zero). By choosing the parameters m, $p$, $k_f$ in equations (22) and (23), different adhesion between cells can be modeled.

**Acknowledgements**

We thanks Z. Nikiticheva (BIN RAN) for providing the anatomical sections of the plant embryos, N. Pakudin and M.Gavrilovich for image comparison analysis. The work of N. Morozova and O.Butuzova was carried out within the framework of the state assignment to Komarov Botanical Institute RAS № AAAA-A18-118051590112-8. The research by A. Minarsky was supported by Ministry of Science and Higher Education of Russian Federation (assignment 1.9788.2017/BCh). The work of N. Bessonov, O.Butuzova, A.Minarskiy, A.Tosenberger was supported by IHES program on mathematical biology (Simons foundation).

32. Nikiticheva Z, Proskurina O (1990) Juncaginaceae. In: Yakovlev MC, editor. Comparative embryology of flowering plants. V. 5. Butomaceae-Lemnaceae. Leningrad: Nauka. pp. 28-34.

**Supplemented materials 1. Morphogenesis Software - Instruction for users**

The program https://github.com/nickbessonov/morphogenesis-article/blob/master/Morphogenesis.zip is launched by the button RUN on the top of the main window of **Morphogenesis Software,** which opens the dialog box "Begin" (Figure 1S).

**Figure 1S.** Dialog panel "Begin"

It allows to set a matrix of a zygote (the initial cell). It means that for each of the eight sectors of the initial cell the amount of the 8 coding substances A, G, F, K, P, D, X, Y in this sector can be assigned, thus providing a coding matrix, associated with this cell. The matrix can be set in 3 following ways:

- each of the 64 coding numbers is manually entered into the fields of matrix elements indicated on the Begin dialog box.

- each coding number is set randomly in the chosen range from 0 to $m_h$ by clicking the button "Random matrix"; the parameters $m_h$ (the maximal amount of coding molecule) and P (a percentage ([0,1]) of heterogeneity in a zygotic matrix, showing the maximal dispersion of its elements) should be set in a corresponding dialog boxes on the panel.

- each coding number is set randomly in the chosen range from 0 to $m_h$, but with a predominance of the values located on the diagonal of the matrix - by clicking the button "Diagonal matrix". The magnitude of the predominance is indicated by the parameter $p$ and is determined by the following rule: if an element lies on the diagonal of the matrix then its value is chosen randomly in the range $[pm_h, m_h]$, if an element does not lie on the diagonal of the matrix, then its value is randomly set in the range $[0, pm_h]$.

The set range from 0 to $m_h$ will be automatically kept for all cells of the embryo.

Next a set of parameters λ, μ, d, $k_3$ (=X) implementing the developmental Rules for Cell events should be specified in the corresponding windows of the panel "Begin".

After setting the matrix and specifying a set of parameters the program can be started by the button "Create Initial Cell", which will create a zygote with a given matrix. The development is started by the button "Run" on the panel "Begin", and the visualization of the development of the organism and of the corresponding graph(tree) occurs based on the given initial matrix and parameters (See Fig.1). On the graph, the edge corresponding to cell event "Growth" is marked with a coefficient of growth (which is the enlargement of cell volume during this cell event), the "Internal cell event" is marked by number "1" on the corresponding edge, and "No cell event" is marked with 0 on the corresponding edge). To stop the development one should press the computer keys "CTRL Q".

The check box "Signaling" at the bottom of the Begin panel allows to switch on the developmental program with signaling (see Section 3.4). Without checking this check box the behavior of cells is modeled as independent on the signaling. The check box "Sectors" allows marking the location of the numerated sectors on the surfaces of the cells, which is important for studying signaling (see Fig. 2C). Also, on this panel there is a choice of a possible response to the middle signaling: cell division, cell growing or no cell event.

The button "Advanced" on the top of the main window opens the dialog panel "Advanced" (Figure 2S), which displays a set of elements of software managing. The most important ones are:

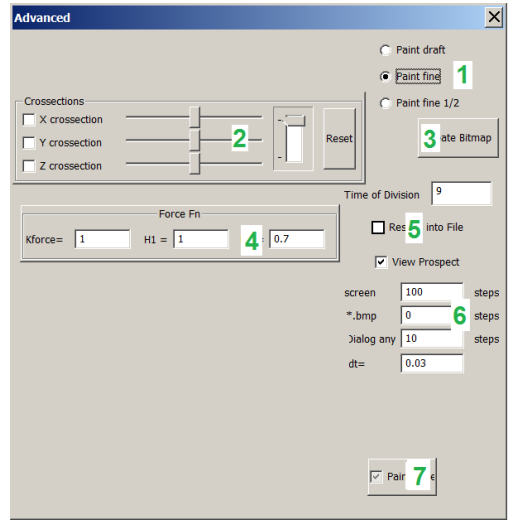

Figure 2S. Dialog Panel Advanced

-the element (shown on Figure 2S, point 2) which allows to make slices (sections) of the embryo perpendicular to one of the axes of coordinates x, y, z (Figure 3S a,b). The position of each slice on the chosen axe can be changed by the corresponding cursor, thus allowing observing all internal structure of the body.

-the element (shown on Figure 2S. point 3) which saves the current image of the main window (embryo, developmental tree, initial matrix) in the bmp file.

-the element (shown on Figure 2S. point 7) which allows displaying a tree of cell events alongside to a growing embryo (see Figure 1).

The remaining elements on this dialog panel (not mentioned here) are currently set by default and not required in the regular work with the program.

Any cell of an embryo visible on the screen can be marked by a mouse click, and simultaneously the chosen cell and its path from the zygote will be shown on the corresponding graph (tree) as a fat line (Figure 3S, c,d) For this, the check box "Selected cell" should be marked at the Begin panel. And vice versa, by the clicking on any of free vertices of the graph detects and marks a corresponding cell in the embryo. For marking or detecting an internal cell it is necessary to make corresponding sections.

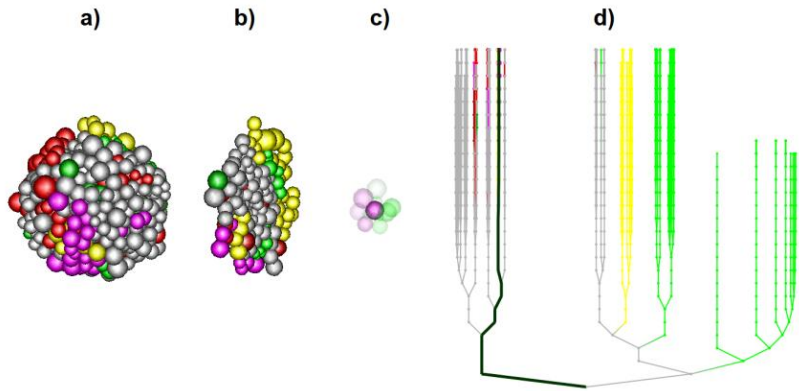

**Figure 3S. Examples of a software managing**. a) growing embryo; b) the cross-section of an embryo; c) a selected cell (bright colored) surrounded by its neighbors (faint colored); d) corresponding developmental tree.

The epigenetic matrix of this cell can be seen by clicking the button " Selected Cell " on the top of main window opening the dialog panel "Selected Cell" (Figure 4S). The matrix of a chosen cell is presented together with two additional matrixes $â_a$ and $â_d$, important for studying signaling (see Section 3.4). There is also an option to see a chosen cell with a set of its neighboring cells (Figure 3S,c), specifying a desirable neighborhood by the "area" buttons on the dialog panel "Selected Cell".

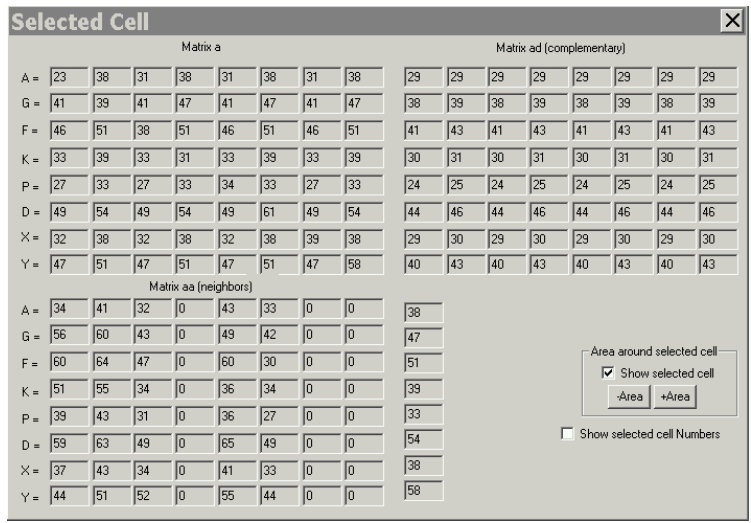

**Figure 4S. Dialog panel "Selected Cell".** Left upper table is a matrix (a) of a selected cell, left bottom table is an "adjacent matrix" ($â_a$), which is combined from the corresponding adjacent sectors of the neighbor cells; the upper right table is a "complementary to adjacent" matrix ($â_d$). Zero in the "adjacent matrix" means the absence of a neighbor cell adjacent to a given sector of a cell.

**Supplemented materials 2. Algorithm for trees comparison**

Let us consider a set of cell events $C = \{\emptyset, I, G, X, Y, Z, A\}$, with elements: null element ($\emptyset$), identity (no change), growth, divisions in x, y and z directions respectively, and apoptosis.

***Def:*** **Tree of cell events** $T = (V, E, e)$, where $V$ is a set of vertices (cell events), $E \subseteq V \times V$ is a set of directed edges (cells states - epigenetic code...), and $e: V \to C$ a mapping from vertices of the tree $T$ to the set of cell events $C$, such that $(V, E)$ is a binary tree.

The **level** $l(v)$ of a vertex $v \in V$ is the length of the path from the root of the tree to the vertex. The maximal level of the tree we denote with $l_{max}$.

**Remark:** The choice of cell events being represented by vertices is not the best one, but at this stage it is easier to define cell division as a single element of a graph (a vertex) instead of having two elements (two edges). Hence, for simplicity, in this description cell events are represented by vertices, and later this can be changed.

***Def:*** If $T = (V, E, e)$ is a tree of cell events, we define the **parent function** which gives the $n$-th ancestor of a node as $p: V \times \mathbb{N}_0 \to V$ such that $\forall a \in V$ and $\forall n \in \{0, \ldots, l(v)\}$ there $\exists b_0, b_1, \ldots, b_n \in V$ such that $(b_{i-1}, b_i) \in E$ $\forall i \in \{1, \ldots, n\}$. In case when $n > l(v)$, we define $p(v, n) = \emptyset$. Also we define $p(v, 0) = v$.

***Def:*** If $T = (V, E, e)$ is a tree of cell events for any two vertices $v, w \in V$ we define a **path** $\langle v, w \rangle$ between them as the ordered n-tuple $(v = v_0, v_1, \ldots, v_n = w)$ such that $\forall i \in \{1, \ldots, n\}$ is $(v_{i-1}, v_i) \in E$, and that $\forall i, j \in \{1, \ldots, n\}, \; i \neq j$ is $v_i \neq v_j$. $n$ is then the path length.

In any tree of cell events $T(V, E, e)$ for every two vertices there exists a path between them and that path is unique.

***Def:*** If $T = (V, E, e)$ is a tree of cell events, for every cell event $c \in C$ we define its corresponding **cell event δ-function** $\delta_c: V \to C$ as

$$\delta_c(v) = \begin{cases} 1, & e(v) = c \\ 0, & e(v) \neq c \end{cases}, \quad v \in V \tag{S1}$$

Hence we have 6 different cell event δ-functions ($\delta_X$, $\delta_Y$, etc.).

***Def:*** If $T = (V, E, e)$ is a tree of cell events, for every cell event $c \in C$ we define its corresponding **cell event measure** $m_c: P \times V \to \mathbb{N}_0$

$$m_c(v, w) = \sum_{i=0}^{n} \delta_c(v_i), \tag{S2}$$

where $\langle v, w \rangle = (v = v_0, v_1, \ldots, v_n = w) \in P_T$ is the path between the two vertices, and $P_T$ is the set of all paths of the tree $T$. Hence we have 6 different cell event measures ($m_X$, etc.).

***Def:*** For all leaves $v_1, \ldots, v_k$ of level $n$ of the tree $T = (V, E, e)$ we define:

- The **sum** $m_c^n$ of measures of all paths from the root of the tree to the nodes $v_1, \ldots, v_k$ as:

$$m_c^n = \sum_{i=1}^{k} m_c(v_i). \tag{S3}$$

- The **mean** $\bar{m}_c^n$ of measures of all paths from the root of the tree to the nodes $v_1, \ldots, v_k$ as:

$$\bar{m}_c^n = \frac{1}{k}\sum_{i=1}^{k} m_c(v_i). \tag{S4}$$

- The **standard deviation** $\sigma_c^n$ of measures of all paths from the root of the tree to the nodes $v_1, \ldots, v_k$ as:

$$\sigma_c^n = \sqrt{\frac{\sum_{i=1}^{k}(m_c(v_i) - \bar{m}_c^n)^2}{k}}. \tag{S5}$$

- The **normalized standard deviation** $s_c^n$ of measures of all paths from the root of the tree to the nodes $v_1, \ldots, v_k$ as:

$$s_c^n = \frac{\sigma_c^n}{\bar{m}_c^n \sqrt{k-1}}. \tag{S6}$$

If $\bar{m}_c^n = 0$, then we define $s_c^n$ as 1.

***Def:*** For two trees $T_1 = (V_1, E_1, e_1)$ and $T_2 = (V_2, E_2, e_2)$ we define the **difference between trees** as:

$$\Delta(T_1, T_2) = 1 - \frac{1}{n+1}\sum_{i=0}^{n} \Delta^n, \tag{S7}$$

where $\Delta^n$ is given by

$$\Delta^n(T_1, T_2) = \prod_{c \in C} \left(1 - \frac{|m_c^n(T_1) - m_c^n(T_2)|}{m_c^n(T_1) + m_c^n(T_2)}\right)\left(1 - \left(s_c^n(T_1) - s_c^n(T_2)\right)\right). \tag{S8}$$

If for $m_c^n(T_1) + m_c^n(T_2) = 0$, then we take the part of the product for $c \in C$ in equation (S8) as 1.

The similarity between trees is then 1-$\Delta(T_1, T_2)$.

**3 examples of two trees comparison (shown on Figure 5S).**

P1:

- Uneven number of branches during development leading to the similar final outcome.

- Cell events inverted in time (at level 1 and level 2).
- Similarity score by levels: L0 - 0%, L1 - 67%, L2 - 100%
- Total similarity score of development: 56%

P2:

- Inversion of branches (left and right) at level 2.
- Left and right side inverted.
- Similarity score by levels: L0 - 100%, L1 - 100%, L2 - 100%
- Total similarity score of development: 100%

P3:

- Inversion at level 2, but level 3 is not inverted (does not follow the inversion made at level 2).
- Similarity score by levels: L0 - 100%, L1 - 100%, L2 - 93%
- Total similarity score of development: 98%

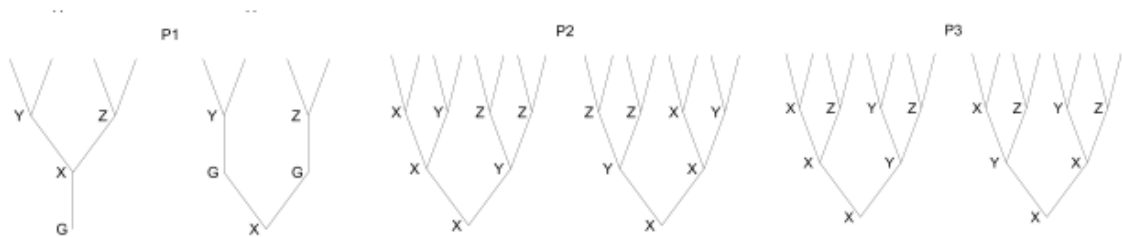

**Figure 5S. 3 pairs of trees, compared by the algorithm in the examples P1,P2,P3.**

**Supplemented materials 3. The best zygotic matrices found for the developmental trees, most close to the actual ones, and the corresponding trees.**

The presented integers should be read row by row (8 integers in each) to provide a matrix.

**Myriophyllum spicatum**

Score:12;1.00;1.00;1.00;0.95;0.95;0.94;0.93;0.92;0.90;0.89;0.88;0.87;

InitMatrix:69;7;66;58;43;36;31;67;46;55;33;31;90;25;91;81;20;80;6;99;81;3;35;70;44;95;41;81;62;96;24;35;28;81;82;57;46;93;88;43;96;14;26;48;90;6;25;78;51;14;25;99;72;4;8;29;44;68;98;82;81;8;11;46

**Polygala major**

Score:12;1.00;0.78;0.80;0.79;0.79;0.80;0.79;0.79;0.79;0.80;0.81;0.81;

InitMatrix:45;23;6;65;99;4;71;96;85;80;29;17;29;61;97;13;89;9;36;45;93;3;77;63;87;74;11;66;27;53;26;37;51;42;25;40;43;19;46;27;80;44;37;50;83;32;91;98;45;88;20;27;83;16;15;96;48;6;79;28;52;25;98;42

**Triglochin palustris**

Score:12;1.00;1.00;0.98;0.92;0.88;0.87;0.84;0.84;0.83;0.83;0.82;0.82;

InitMatrix :72;29;97;46;44;37;51;62;55;90;94;19;86;25;93;10;17;30;7;90;94;80;92;74;20;42;45;62;34;2;96;61;32;5;8;4;69;2;41;10;54;47;97;56;44;54;96;67;41;72;25;5;52;37;83;17;23;32;56;2;24;11;43;2

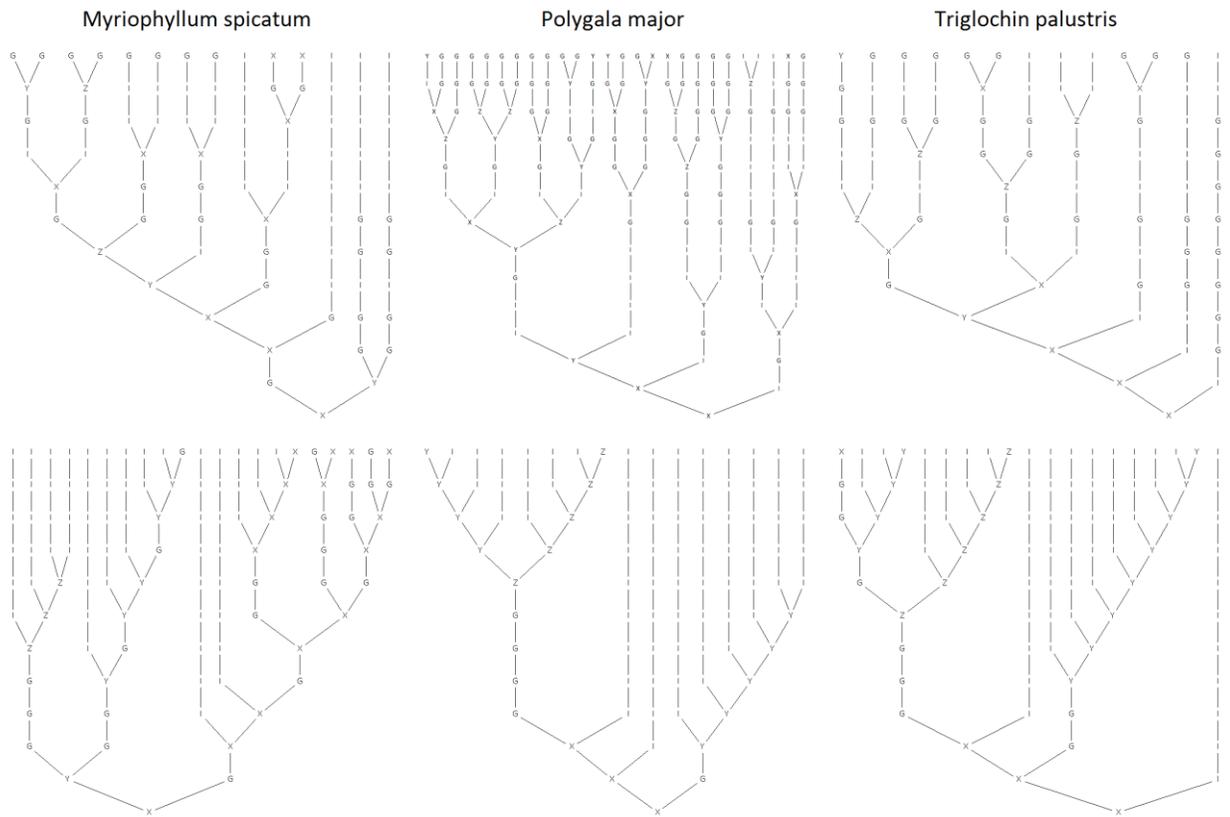